\documentclass{osa-article}

\journal{osajournal}

\articletype{Research Article}

\begin{document}

\title{Photonic sampled and quantized analog-to-\\digital converters on thin-film lithium niobate platform}

\author{Donghe Tu,\authormark{1,2} Xingrui Huang,\authormark{1,2} Yang Liu,\authormark{1,2} Zhiguo Yu,\authormark{1} and Zhiyong Li\authormark{1,*}}

\address{\authormark{1}State Key Laboratory on Integrated Optoelectronics, Institute of Semiconductors, Chinese Academy of Sciences, Beijing 100083, China\\
\authormark{2}College of Materials Science and Opto-Electronic Technology, University of Chinese Academy of Sciences, Beijing 100083, China}
\email{\authormark{*}lizhy@semi.ac.cn} 



\begin{abstract}
In this paper, an on-chip photonic sampled and quantized analog-to-digital converter (ADC) on thin-film lithium niobate platform is experimentally demonstrated. Using two phase modulators as a sampler and a 5$\times$5 multimode interference (MMI) coupler as a quantizer, an 1 GHz sinusoidal analog input signal was successfully converted to a digitized output with a 20 GSample/s sampling rate. To evaluate the system performance, the quantization curves together with the transfer function of the ADC were measured. The experimental effective number of bits (ENOB) was 3.17 bit. The demonstrated device is capable of operating at a high frequency up to 70 GHz, making it a promising solution for on-chip ultra-high speed analog-to-digital conversion. 
\end{abstract}

\section{Introduction}
The analog-to-digital converter (ADC), which converts analog signals to digital signals, plays an important role in modern advanced information systems, such as radars, high speed communication systems, and precision measuring instruments. Nevertheless, conventional electronic ADCs are not suitable for ultra-high speed signal processing and large bandwidth applications since the serious aperture jitter and limited comparator bandwidth \cite{RN1}. To overcome these bottlenecks, optical ADCs which utilize photonic components to implement the conversion were proposed and have become a hotspot in the field of optoelectronic technologies \cite{RN2}.

In recent years, many optical ADC schemes have been proposed and demonstrated, which can be divided into four major types \cite{RN1}: photonic assisted (PA) ADC \cite{RN3,RN4}, photonic sampled (PS) ADC \cite{RN2,RN5,RN6}, photonic quantized (PQ) ADC \cite{RN7} and photonic sampled and quantized (PSQ) ADC \cite{RN8,RN9,RN10,RN11,RN12,RN13,RN14,RN15,RN16,RN17,RN27}. PA, PS and PQ ADCs still require several high speed electronic ADCs for either sampling or quantizing, resulting in additional power consumption and system complexity. The PSQ scheme, which performs sampling and quantization both in the optical domain, can avoid these problems and has attracted much attention in recent years \cite{RN8,RN9,RN10,RN11,RN12,RN13,RN14,RN15,RN16,RN17,RN27}. PSQ ADC can be further subdivided into two broad categories by different quantization methods: optical amplitude quantization (OAQ) schemes \cite{RN10,RN11,RN12,RN13} and phase-shifted optical quantization (PSOQ) schemes \cite{RN8,RN9,RN14,RN15,RN16,RN17,RN27}. In OAQ schemes, different optical amplitude was distinguished via high optical nonlinearities . For example, Ref. \cite{RN13} reported a 2 bit all optical ADC using photonic crystal based cavities. In \cite{RN10}, an optical ADC with a sampling rate of 40 GS/s and the effective number of bits (ENOB) of 3.79 was demonstrated by using a highly non-linear fiber to induce soliton self-frequency shift. However, OAQ schemes usually require high input optical power and a long interactive length, which lead to additional power consumption and difficulty of chip level integration. As for PSOQ schemes, the quantization is accomplished by the interference of the lights with different phases. The interference can be achieved by on-chip Mach-Zehnder interferometers, and the requirement of high optical nonlinearities in OAS schemes is avoided. Thus, the optical power could be more efficient and the footprint could be relatively small. In Ref. \cite{RN9}, an optical ADC with sampling rate of 40 GS/s and ENOB of 3.45 using a phase shifter array was reported. In Ref. \cite{RN14,RN15}, an optical analog-to-digital conversion scheme using Mach–Zehnder modulators (MZM) with identical half-wave voltages was reported. Compared with the OAQ schemes, the above-mentioned PSOQ schemes effectively reduced the input optical power, but they still used discrete devices. In order to achieve an on-chip integration and further downsize the footprint of the system, a quantization approach using a 3$\times$5 step-size  multimode interference (MMI) coupler was proposed \cite{RN16}. The scheme in Ref. \cite{RN16} was further investigated in Ref. \cite{RN17}. A quantizer utilizing integrated photonics with ENOB of 3.31 was fabricated on SOI platform. However, the authors in \cite{RN17} only focused on the quantizer, which is only one part of the PSQ ADC. The sampling process and the complete analog-to-digital conversion were not engaged. Furthermore, sampling a high speed analog signal for PSOQ schemes on SOI platform is challenging due to three main drawbacks. Firstly, the large half wave voltage ($V_\pi$) of high speed SOI phase modulators makes it difficult to meet the 2$\pi$ phase shift requirement in PSOQ schemes. The electrical divers that can provide enough output power are hard to achieve, and a breakdown may occur when using reverse biased PN junction modulators. Next, the nonlinearities of the SOI modulators would lead to sampling errors. The modulation efficiency decreases when the input voltage increasing because of the nonlinearities of the plasma dispersion effect. Finally, the relatively small electro-optical (EO) 3-dB bandwidth of SOI modulators limits the sampling bandwidth of the ADC, which makes it difficult to competent ultra-fast applications.

To overcome these above-mentioned shortcomings, an on chip PSQ ADC on thin-film lithium niobate (TFLN) platform was demonstrated in this paper. The Pockels effect of lithium niobate provides a linear modulation on TFLN platform. And over the past 4 years, many high speed and low $V_\pi$ electro-optical modulators on TFLN platform are reported \cite{RN18,RN19,RN20}. In \cite{RN18}, a EO 3-dB bandwidth of a 80 GHz and $V_\pi$ of 2.3 V modulator was demonstrated. It showed the capability of TFLN platform for large bandwidth sampling and meeting the 2$\pi$ phase shift requirement for the MMI quantizer with a lower drive voltage.

In our work, a PSQ ADC on a TFLN chip consisting of two beyond 70 GHz EO 3-dB bandwidth phase modulators and a 5$\times$5 MMI was demonstrated. The input analog signal was sampled by an optical pulse train using the two phase modulators. An optical 72$^{\circ}$ hybrid based on the 5$\times$5 MMI proposed in our previous study \cite{RN21} was used for a 10-level quantization. The quantization results were then coded and converted to a digital output with 10 digitized levels. To evaluate the system performance, the quantization curves and the transfer function of our proposed ADC were measured and analyzed. And an 1 GHz electrical analog sinusoidal signal was converted to a digital output with a 20 GS/s sampling rate. To the best of our knowledge, this is the first time that the sampling and the quantization were demonstrated on a single chip for PSQ ADCs. The ENOB was measured to be 3.17 and the input sampling EO 3-dB bandwidth was beyond 70 GHz. Therefore, our work has provided a potential solution towards high-bandwidth on-chip analog-to-digital conversions and optical signal processing systems.

\section{Methods}
\begin{figure}[h!]
\centering
\includegraphics[width=12cm]{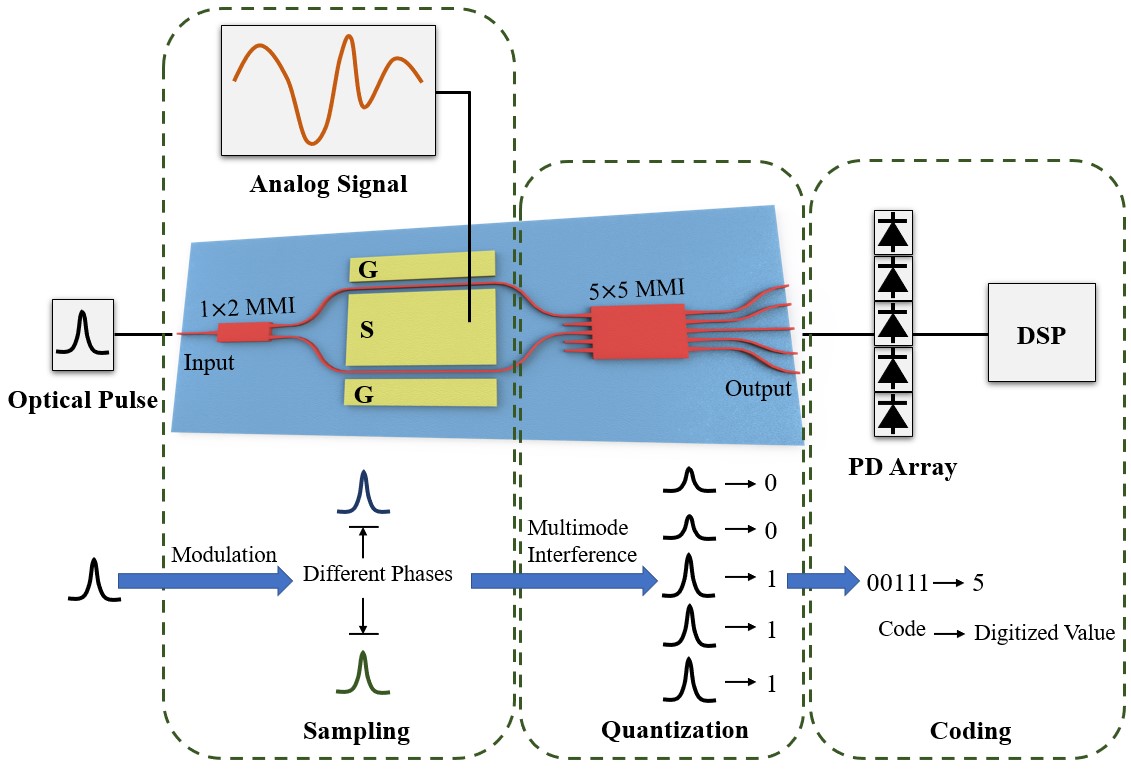}
\caption{Conceptual diagram of our proposed PSQ ADC}
\label{fig:1}
\end{figure}

A conceptual diagram of our proposed PSQ ADC is shown in Fig. \ref{fig:1}. An analog signal is sampled by an optical pulse train using two identical 1-cm long phase modulators with opposite modulation phase shift. Next, the two sampled lights are quantized by a 5$\times$5 MMI. Following by a photodetector (PD) array and a digital signal processing (DSP) unit, the signal is then coded and transferred to digitized values. The following contents in this section will introduce the operation principles of sampling and quantization, which are the two main functions of our proposed ADC.

\subsection{Sampling}

As shown in Fig. \ref{fig:1}, an optical pulse train is fed into the TFLN ADC chip as an optical source. The optical pulse train can be represented as

\begin{equation}
E_{\textup{input}} = A\cdot e^{-j\omega t} \cdot f(t)
\label{eq1}
\end{equation}   
where $A, \omega , f(t)$ represent the amplitude, angular frequency and the normalized envelope of the pulse shape, respectively. To simplify the calculation, the initial phase of the optical pulse is considered to be zero. After the input analog signal is phase modulated, the amplitude of the two output lights ($E_0$, $E_1$) can be expressed as

\begin{equation}
\begin{split}
E_{\textup{1}} = \frac{\sqrt{2}}{2} A\cdot e^{-j\omega t}\cdot e^{-j(\phi_1 + \frac{1}{2} \Delta\phi_{\textup{signal}})} \cdot f(t)\\
E_{\textup{2}} = \frac{\sqrt{2}}{2} A\cdot e^{-j\omega t}\cdot e^{-j(\phi_2 - \frac{1}{2} \Delta\phi_{\textup{signal}})} \cdot f(t)
\label{eq2}
\end{split}
\end{equation}   
where $\phi_1$ and $\phi_2$ are the random phase shifts induced by the non-uniformity of the waveguides of the two phase modulators due to fabrication errors. $\frac{1}{2} \phi_{\textup{signal}}$ represent the amount of change in phase by both phase modulation. Since the modulation on TFLN platform utilizes the Pockels effect, the phase change ($\frac{1}{2} \phi_{\textup{signal}}$) is proportional to the amplitude of the input analog signal, which can be expressed as

\begin{equation}
\frac{1}{2} \Delta\phi_{\textup{signal}}=\frac{\pi V_{\textup{analog}}}{V_\pi}
\label{eq3}
\end{equation}  
where $V_{\textup{analog}}$ is the amplitude of the input analog signal and $V_\pi$ is the half wave voltage of the phase modulators. The phase difference between the two modulated lights  ($\Delta \phi$) can be written as

\begin{equation}
\Delta \phi=\phi_1-\phi_2+\Delta\phi_{\textup{signal}}
\label{eq4}
\end{equation}  
where $\phi_1-\phi_2$ is a constant. According to Equation (\ref{eq3}) and (\ref{eq4}), the sampling process induces an additional phase difference ($\Delta \phi$) between the two modulated lights, which increases linearly with the amplitude of the input analog signal.

\subsection{Quantization}

\begin{figure}[b]
\centering
\includegraphics[width=9cm]{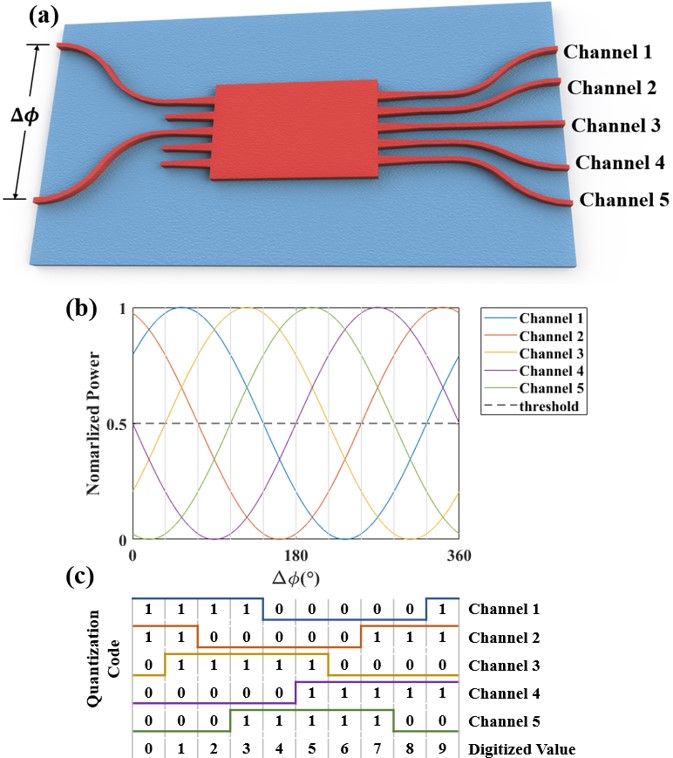}
\caption{(a) Schematic diagram of the optical 72$^{\circ}$ hybrid. (b) Quantization curves of the quantizer. (c) The corresponding quantization codes for ten quantization levels }
\label{fig:2}
\end{figure}

The quantization process in our proposed ADC converts the two modulated lights with different phases into 5-channel outputs with 10 quantization levels. As shown in Fig. \ref{fig:2} (a), after the analog signal is sampled, the two lights with a phase difference ($\Delta \phi$) are fed into a 15-um wide and 235-um long 5$\times$5 MMI. According to our previous study \cite{RN21}, when the input channels are the first and the third ports, the 5$\times$5 MMI can be used as an optical 72$^{\circ}$ hybrid with an acceptable phase deviation in a relatively large wavelength range. The output optical power intensities of the five channels of the MMI ($I_1 \sim I_5$) can be successively expressed as

\begin{equation}
\begin{aligned}
I_1&=\frac{1}{5} A^2 [1+cos(\Delta \phi-\frac{2}{5}\pi)]\\
I_2&=\frac{1}{5} A^2 [1+cos(\Delta \phi)]\\
I_3&=\frac{1}{5} A^2 [1+cos(\Delta \phi-\frac{4}{5}\pi)]\\
I_4&=\frac{1}{5} A^2 [1+cos(\Delta \phi+\frac{2}{5}\pi)]\\
I_5&=\frac{1}{5} A^2 [1+cos(\Delta \phi+\frac{4}{5}\pi)]
\label{eq5}
\end{aligned}
\end{equation}  

Equation (\ref{eq5}) shows that the optical power intensity of each channel is sinusoidally related to $\Delta \phi$. As one can see in Fig. \ref{fig:2} (b), a 72$^{\circ}$ phase shift is induced between the adjacent quantization curves. Next, a threshold power is set at the middle of the normalized power to determine the digital output of each channel. For instance, if the output power intensity of one channel is above the threshold, it is decided to be 1. Otherwise is decided to be 0. The five transmission curves divide the threshold line into 10 regions, corresponding to 10 quantization levels. In Fig. \ref{fig:2} (c), 10 groups of codes for the 10 quatization levels are shown. According to $log_210=3.32$, the above analysis indicates that the theoretically quantization bit of the 5$\times$5 MMI is 3.32.

\section{Device fabrication and experimental results}
In this section, the fabrication process of the proposed optical ADC on the TFLN platform was illustrated. Then, the measurements of the quantization curves and the transfer function of the proposed ADC were introduced, and the equivalent ENOB was obtained. Finally, the experiment of digitizing an 1 GHz sinusoidal signal with a 20 GS/s optical pulse train was demonstrated.

\subsection{Device fabrication}
\begin{figure}[h!]
\centering
\includegraphics[width=12cm]{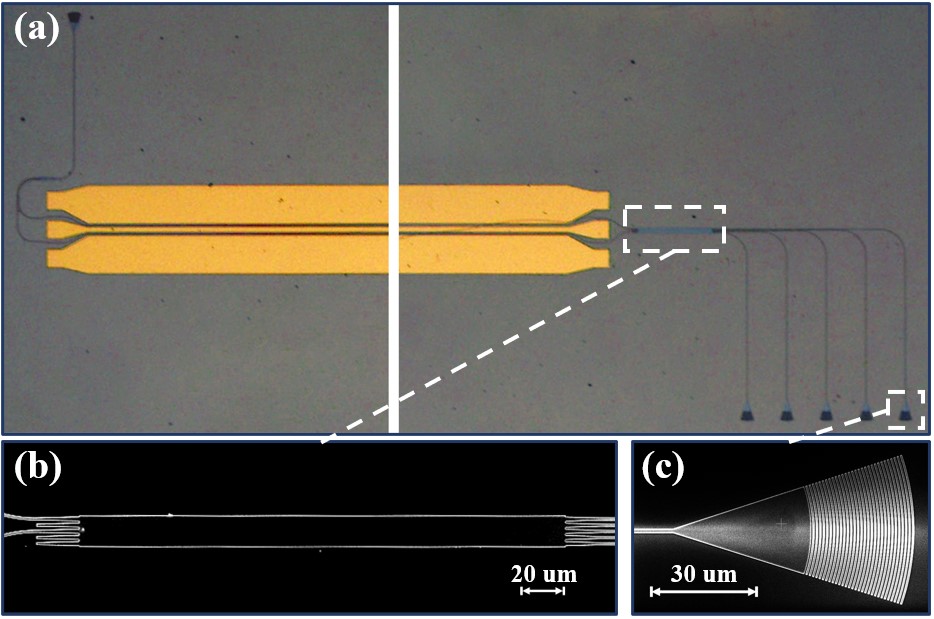}
\caption{(a) Microscope image of the demonstrated ADC. SEM image of the fabricated 5$\times$5 MMI (b) and the grating coupler (c).}
\label{fig:3}
\end{figure}

Fig. \ref{fig:3}(a) shows the microscope image of the demonstrated ADC, and Fig. \ref{fig:3}(b) and (c) show the scanning electron microscope (SEM) image of the 5$\times$5 MMI and the grating coupler. The detailed fabrication process of the devices can be found in our previous work \cite{RN23}. The devices were fabricated on a x-cut lithium niobate-on-insulator with a 400-nm thick TFLN and a 3-$\mu$m buried oxide. Firstly, the wafer was cleaned by a piranha solution (H$_2$SO$_4$/H$_2$O$_2$ = 3:1, 80$^{\circ}$C) and an oxygen plasma. Next, a microlithography adhesion promoter named Surpass 3000 and e-beam hydrogen silsesquioxane (HSQ) were sequentially spin-coated to the surface at 3000 rpm for 60s. After a 80$^{\circ}$C, 4 min pre-bake, the e-beam lithography was performed by the JBX-8100FS system with an accelerating voltage of 100 KV. Followed by the resist development, the TFLN was partially dry-etched by 200 nm using argon-based inductively coupled plasma reactive ion etching with approximately 1:2.5 etch selectivity between TFLN and HSQ. Finally, 1-$\mu$m thick gold electrodes were fabricated using UV lithography and a standard lift-off process. The total area of our demonstrated ADC (including the grating couplers) is approximately 0.33$\times$11.8 mm$^2$.

\subsection{Measurement of quantization curves and tranfer function}

\begin{figure}[h!]
\centering
\includegraphics[width=13cm]{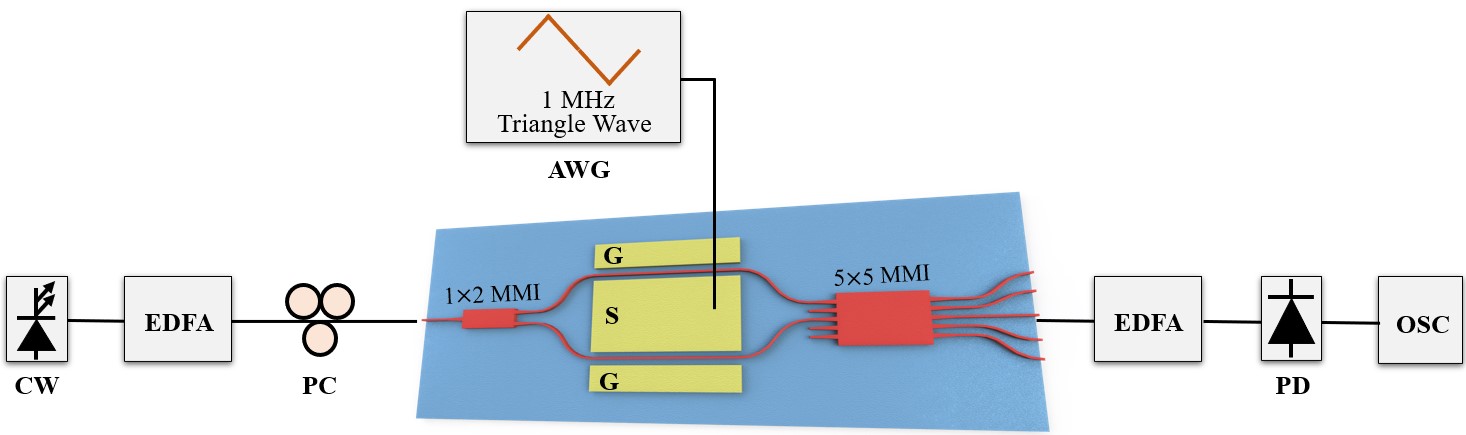}
\caption{Experimental setup of measuring the quantization curves and transfer function}
\label{fig:4}
\end{figure}

Fig. \ref{fig:4} shows the experimental setup of the quantization curves and transfer function measurements. The ENOB can then be obtained from the experimental results. The input light was generated by a continuous wave (CW) laser. It was amplified by an erbium-doped fiber amplifier (EDFA) to 20 dBm and then connected to a polarization controller (PC). Next, it was fed into the ADC chip using a grating coupler and split into two lights by an 1$\times$2 MMI. After oppositely modulated by an 1 MHz triangular signal generated from an arbitrary waveform generator (AWG), the two modulated lights were connected to the input channels of the 5$\times$5 MMI. Therefore, the phase difference between the two lights ($\Delta \phi$) changed linearly over the half period of the triangular signal. After passing through the 5$\times$5 MMI, the output lights were amplified by another EDFA to 3 dBm and then transferred to electrical signals by a PD with a responsibility of 0.7 A/W. The signals were recorded by an oscilloscope (OSC) and processed into quantization curves by a linear time-phase mapping.

\begin{figure}[t!]
\centering
\includegraphics[width=13cm]{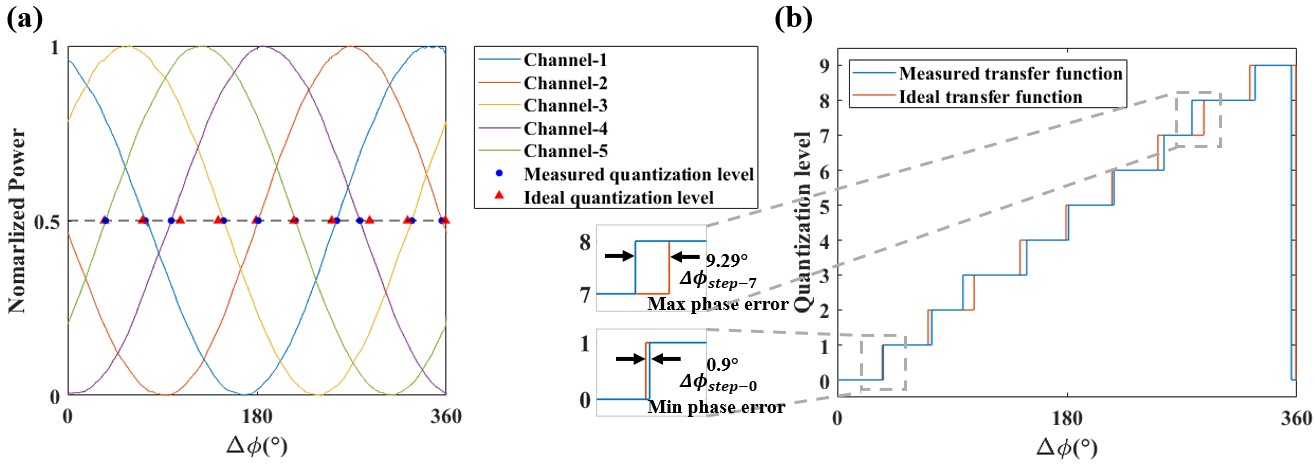}
\caption{(a) Normalized quantization curves. (b) Transfer function of the optical ADC and the ideal value. Insets: the maximum and the minimum phase errors.}
\label{fig:5}
\end{figure}

The normalized quantization curves of the 5 channels are shown in Fig. \ref{fig:5}(a). The measured curves were smoothed to reduce the white noise induced by the PD. A decision threshold represented as the black dash line are set at the middle of the normalized power to determine the 10 quantization levels mentioned above. The dots are the intersections of the curves and the threshold, which represent the measured phase of the 10 quantization levels. The ideal phases of the 10 quantization levels are marked with triangles. Fig. \ref{fig:5}(b) shows the transfer function of our proposed ADC. The transfer function is obtained based on the quantization curves. The blue line is the measured value while the red line is the ideal one. The maximum and minimum phase errors of the quantization levels,${\Delta\phi_{step-7}}$ and ${\Delta\phi_{step-0}}$, are shown in the upper and the lower insets respectively. ${\Delta\phi_{step-7}}$ is 9.29$^{\circ}$ while ${\Delta\phi_{step-0}}$ is 0.9$^{\circ}$, which means the maximum relative error is 12.9$\%$ and the minimum one is 1.25$\%$. The equivalent ENOB can be calculated using Equation (\ref{eq6}) according to Ref. \cite{RN22}.

\begin{equation}
ENOB=\frac{20}{6.02}log_{10}\frac{P_{\textup{FS}}/\sqrt{12}}{\sqrt{\frac{1}{12}(\frac{P_{\textup{FS}}}{Q})^2+\frac{1}{Q}\sum_{i=0}^{Q-1}{\Delta\phi_{step-i}^2}}}
\label{eq6}
\end{equation}  
where $\Delta\phi_{step-i}$ represents the phase error of each quantization level. $P_{\textup{FS}}$ is the maximum value of the phase difference ${\Delta\phi}$. Q is the total number of the quantization levels. Thus, the ENOB was calculated to be 3.17.

The decrease between the theoretically quantization bit of 3.32 and the ENOB of 3.17 is induced by the phase errors mentioned above, which is mainly cause by the inherent phase deviation of a 5$\times$5 MMI, the fabrication errors and the PD noise. The asymmetric input channels lead to a inherent phase deviation of the 5$\times$5 MMI. A 5$\times$5 MMI with input ports of channel 2 and 4 can reduce the inherent phase deviation at a specific wavelength according to Ref. \cite{RN21}. However, it would limit the operating wavelength range and lessen the fabrication tolerances. In order to lower the impact of the PD noise, larger output optical power is needed. One can design and manufacture MMI and phase modulators with lower insertion losses. Also, grating couplers can be replaced by edge couplers to reduce the coupling loss. To reduce the impact of fabrication errors, a compact MMI with sub-wavelength gratings \cite{RN26} can be used to mitigate the effects of etching and substrate non-uniformity. Additionally, an optimized 5$\times$5 MMI design, such as using sub-wavelength structures in Ref. \cite{RN25}, can further decrease the phase errors, which is the research direction in our future works.

\subsection{Analog to digital conversion experiment}
\begin{figure}[h!]
\centering
\includegraphics[width=13cm]{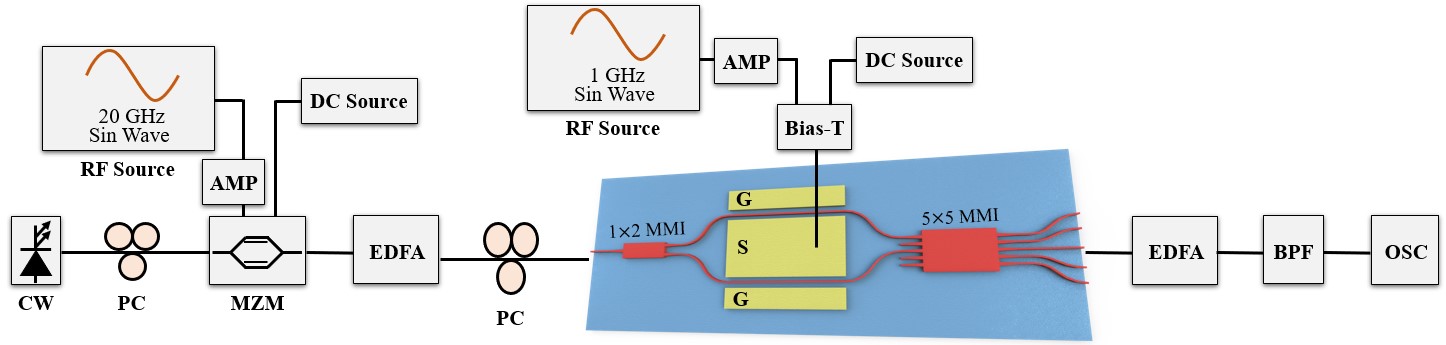}
\caption{Experimental setup for digitizing an 1 GHz sinusoidal electrical signal.}
\label{fig:6}
\end{figure}
Using our presented optical ADC, an 1 GHz sinusoidal electrical analog signal was sampled and quantized. The experimental setup is shown in Fig. \ref{fig:6}. Light generated from a CW was modulated by a commercial MZM. The amplitude of the input 20 GHz analog signal was 0.51 times of the $V_\pi$ of the commercial MZM, and the bias point induced by a DC source was $3/4V_\pi$. Thus, a 20 GS/s optical pulse train with a duty cycle of 37$\%$ was generated. After amplified by an EDFA to the peak power of 20 dBm and connected to a PC, the pulse train was fed into the ADC chip. An 1 GHz sinusoidal electrical analog signal was sampled by the pulse train and then quantized by the 5$\times$5 MMI. The 5 outputs were amplified by another EDFA to the maximum peak power of 3 dBm, and then filtered by a band pass filter (BPF). Finally, an OSC with optical ports was used to recorded the 5 output waveforms.

\begin{figure}[t]
\centering
\includegraphics[width=13cm]{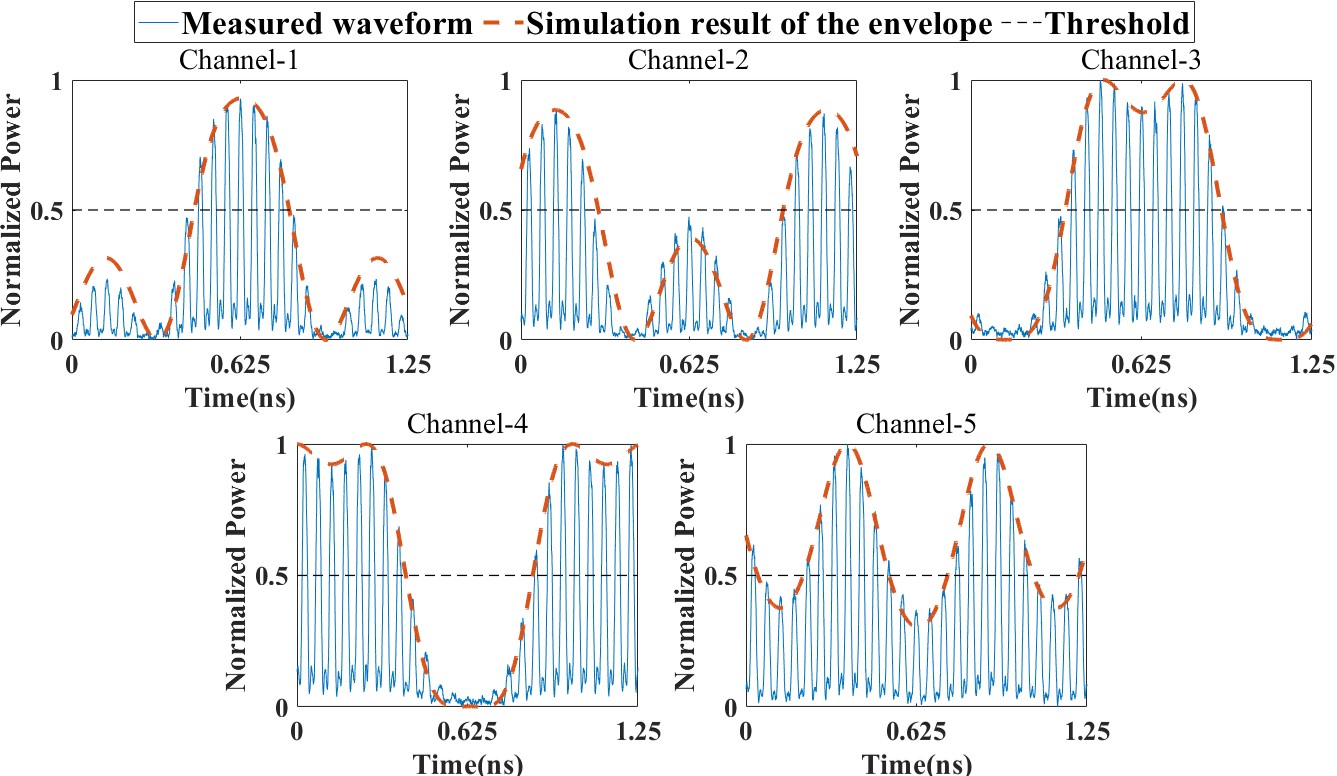}
\caption{Normalized waveforms of the 5 output channels and ideal simulated envelopes.}
\label{fig:7}
\end{figure}

\begin{figure}[h]
\centering
\includegraphics[width=7cm]{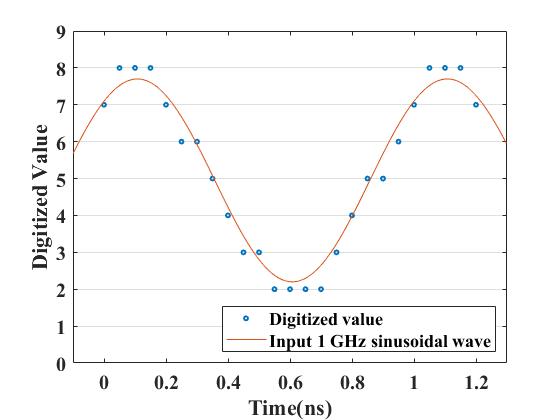}
\caption{Digitized values and the input 1 GHz sinusoidal wave.}
\label{fig:8}
\end{figure}

The blue lines in Fig. \ref{fig:7} show the normalized recorded waveforms of the 5 output channels. The red dash lines are the ideal simulated envelopes of the pulses. The simulation results were based on Equation (\ref{eq5}), where the power was normalized and $\Delta \phi = (V_\pi /\pi)V_{pp}cos(\omega_mt)+\phi_1-\phi_2$. $V_\pi$ is the half wave voltage of the phase modulator. $V_{pp}$ and $\omega_m$ are the peak-to-peak voltage and angular frequency of input 1 GHz analog signal. $\phi_1-\phi_2$ is the initial phase difference mentioned above that can be adjusted by a DC source. The measured waveforms are generally consistent with the simulation results, and the mismatches are mainly due to the PD noise and the phase errors mentioned above. The black dash lines represent the threshold power. For each pulse in the waveforms, if the maximum power is larger than the threshold, the output value of this pulse is decided to be 1, and vice versa. Then, combining the output values of the 5 channels, the quantization codes are obtained. The corresponding digitized values overlaid with the input 1 GHz sinusoidal wave are shown as the circles and the solid line in Fig. \ref{fig:8}. As a result, the analog-to-digital conversion of an 1 GHz sinusoidal wave was successfully implemented. Noted that only 7 of the 10 quantization levels are used in this proof-of-concept experiment, which is due to the $V_{pp}$ of the input analog signal could only be amplified to 0.6 times of the $V_\pi$ of the phase modulators by the amplifier we used. 

\section{Discussion}

\begin{figure}[b]
\centering
\includegraphics[width=8.5cm]{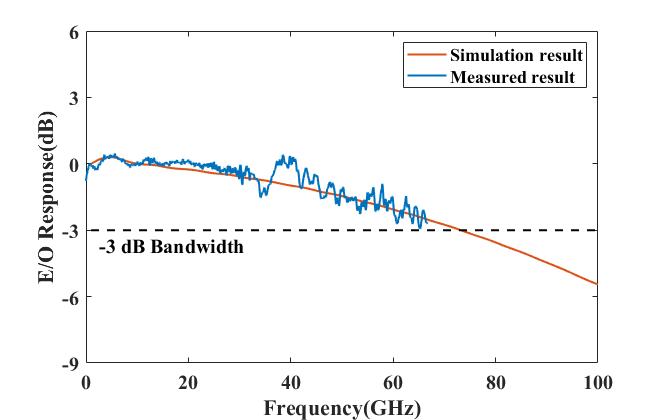}
\caption{Measured and simulated EO response as a function of frequency.}
\label{fig:9}
\end{figure}

\begin{table}[h]
\centering
\caption{\centering \bf Comparison of reported optical ADCs}
\begin{tabular}{m{1cm} m{2cm}<{\centering} m{2.5cm}<{\centering} m{1.5cm}<{\centering} m{2.5cm}<{\centering} m{1.2cm}<{\raggedleft}}
\hline
Year & Type & Sampling 3-dB bandwidth  & Platform & Footprint & Ref.\\ \hline
2011 & PSQ & 20 GHz & Discrete & / &\cite{RN27} \\
2012 & PS & 12 GHz & SOI & 7$\times$3.25 mm$^{\textup{2}}$ &\cite{RN2} \\
2020 & PS & 45 GHz & SOI $^{\textup{a}}$ & 3.3$\times$5 mm$^{\textup{2}}$ &\cite{RN5} \\
2021 & PA & 100 GHz $^{\textup{b}}$ & Discrete & / &\cite{RN4} \\
2021 & PSQ & / & SOI & 7.5$\times$95 $\mu$m$^{\textup{2}}$ $^{\textup{c}}$  &\cite{RN17} \\
\bf{2022} & \bf{PSQ} & \bf{$>$70 GHz} & \bf{TFLN} & \bf{0.33$\times$11.8 mm$^2$} & \bf{This work} \\
\hline
\end{tabular}
\raggedright\footnotesize{$^{\textup{a}}$The filters were on the SOI platform while the modulator for sampling was discrete and off the shelf.}
\raggedright\footnotesize{$^{\textup{b}}$An EO 3-dB bandwith of 100 GHz modulator was used to sample a 320 GHz signal.}
\raggedright\footnotesize{$^{\textup{c}}$The footprint was only for the quantizer while the sampling experiment was not carried out.}
\label{tab1}
\end{table}

The above experimental results show the ENOB of 3.17 and the capability of digitizing analog signals. To further evaluate the ADC performance, we measured the input sampling 3-dB bandwidth. The characterizations of the small signal were performed using a 67 GHz vector network analyzer (VNA). Fig. \ref{fig:9} shows the EO response as a function of frequency. The measured result decays about 2.6 dB from 10 MHz to 67 GHz. The simulated EO response was based on the average modulation voltage model described in \cite{RN24}, which can be expressed as

\begin{equation}
r(f_m)=20log_{10}\frac{\lvert V_{avg}(f_m) \rvert}{\lvert V_{avg}(DC) \rvert}
\label{eq7}
\end{equation}  
where $r(f_m)$ is the EO response of a frequency $f_m$, and $V_{avg}$ is the the average voltage experienced by a photon as it traverses through the modulator, which is given as

\begin{equation}
V_{avg}=\frac{V_g(1+\rho_1)e^{\gamma i\beta_o l}(V_+ +\rho_2 V_-)}{2[e^{\gamma l}+\rho_1 \rho_2 e^{-\gamma l}]}
\label{eq8}
\end{equation} 
where $V_g$ is the amplitude of the driving voltage. $\rho_1 = (Z_c-Z_s)/(Z_c+Z_s)$ and $\rho_2 = (Z_t-Z_s)/(Z_t+Z_s)$ are the input and output reflection coefficients. $V_+$ and $V_-$ are the single pass average voltage due to co-propagation and counter-propagation between microwave and optical wave, where $V_\pm = e^{\pm i \phi_\pm}sin\phi_\pm / \phi_\pm$, and $\phi_\pm = (\beta_e \mp \beta_o)l/2$, l is the length of the phase shifter length.  $\beta_o$ is the complex phase of the optical wave, and $\gamma$ is the propagation constant of microwave. The values of $Z_c$,$\gamma$ and $\beta_o$ used in this simulation were all extracted from experiments.

As shown in Fig. \ref{fig:9}, the simulation result agrees with the measured result, and the EO 3-dB bandwidth is expected to be beyond 70 GHz. This suggests that our presented ADC is capable of operating well with the input frequency up to 70 GHz. In order to accurately sample such high speed analog signals, a pulse source with ultra-narrow pulse width, such as a femtosecond mode-locked laser, should be used to further exploit the performance of the ADC.

A comparison of the sampling 3-dB bandwidth and the footprint for various types of optical ADCs is shown in Table~\ref{tab1}. Compared to those reported optical ADCs, our proposed work has a relatively large sampling 3-dB bandwidth and is on-chip implemented with a compact footprint at the same time, which makes it a potential integrated solution for high-bandwidth analog-to-digital conversions in the field of all-optical signal processing.

\section{Conclusion}
To draw a conclusion, we present an on-chip PSQ ADC on thin-film lithium niobate platform using two phase modulators for sampling and a $5\times 5$MMI for quantization. The quantization curves and the transfer function were measured and analyzed. The ENOB, obtained from the experimental results, was 3.17 bit. Using a 20 GS/s pulse train as an optical source, an 1 GHz input analog signal was converted to a digitized output. The footprint of the ADC is approximately 0.33$\times$11.8 mm$^2$. Furthermore, the demonstrated devices are capable of operating at a high frequency up to 70 GHz. Therefore, the presented ADC provides a promising solution for ultra-high speed analog-to-digital conversions.

\begin{backmatter}
\bmsection{Funding}
This work is supported by the National Key Research and Development Program of China under Grant 2018YFE0201900.

\bmsection{Disclosures}
\noindent The authors declare no conflicts of interest.

\bmsection{Data availability} Data underlying the results presented in this paper are not publicly available at this time but may be obtained from the authors upon reasonable request.

\end{backmatter}








\end{document}